\documentclass[nofootinbib,prd,twocolumn,showpacs,showkeys,preprintnumbers]{revtex4-1}
\usepackage{hyperref,amssymb,amsmath,mathrsfs,bm,graphicx}
\begin{document}
\title {VORTICITY AND ENTROPY PRODUCTION IN TILTED SZEKERES SPACETIMES}
\author{L. Herrera}
\email{laherrera@cantv.net.ve}
\altaffiliation{Also at U.C.V., Caracas}
\author{A. Di Prisco}
\email{adiprisc@fisica.ciens.ucv.ve}
\altaffiliation{Also at U.C.V., Caracas}
\author{J. Ib\'a\~nez  }
\email{j.ibanez@ehu.es}
\affiliation{Departamento de F\'\i sica Te\'orica e Historia de la Ciencia,
Universidad del Pa\'{\i}s Vasco, Bilbao, Spain}
\author{J.Carot }
\email{jcarot@uib.cat}
\affiliation{Departament de  F\'{\i}sica, Universitat Illes Balears, E-07122 Palma de Mallorca, Spain}
\date{\today}
\begin{abstract}
We analyze the properties of the tilted Szekeres spacetime, i.e. the version of such spacetime as seen by a congruence of observers with respect to which the fluid is moving. The imperfect fluid and the kinematical variables associated to the four-velocity of the fluid assigned  by tilted observers are studied in detail. As it happens for the case of the Lemaitre--Tolman--Bondi spacetime, the fluid evolves nonreversibly (with nonvanishing entropy production) and is nongeodesic. However unlike the latter case, the tilted observer  detects vorticity in the congruence of the fluid world lines. Also, as for the nontilted congruence the magnetic part of the Weyl tensor vanishes, reinforcing the nonradiative character of this kind of spacetime. Possible physical implications of these results are discussed.
\end{abstract}
\date{\today}
\maketitle

\section{Introduction}
 In any physical theory, part of the description of a specific scenario is tightly related to the  congruence of observers carrying out the study \cite{HC}.   In the  case of general relativity, the existing arbritariness in the choice of the four velocity in terms of which the energy--momentum tensor is split leads to a variety of physical interpretations for the source of a given spacetime (see \cite{23b}--\cite{38} and references therein).

As it has been shown in the references above, when the two possible interpretations of  a given spacetime correspond to two congruences of observers related by a local Lorentz boost, both the general properties of the source and the kinematical properties of the congruence  would be different.

Particularly  enlightening is the case  of tilted Lemaitre--Tolman--Bondi (LTB)  spacetime whose fluid source  is nongeodesic and  evolves nonreversibly, unlike the nontilted version (see \cite{38} for details).

An important (though obvious) remark is in order at this point: we are assuming general relativity to be valid. This implies that special relativity is always valid at the  local level. Therefore we can always perform a Lorentz boost (locally), thereby ensuring the existence of tilted observers.

In this work we endeavour to study in detail tilted Szekeres spacetimes \cite{1,2}. The motivations  to undertake such a task are many.

 Szekeres dust models have no Killing vectors and therefore represent an interesting generalization of LTB spacetimes\cite{3, 5}.  They have been extensively used as cosmological models (see \cite{4, 6} and references therein) and  in the discussion  of the gravitational arrow of time \cite{7}.

Physical and topological properties of  Szekeres spacetimes, as well as the problem of structure formation  and other cosmological related issues  have been considered in  \cite{7bis, 8, 9, 10, 11, 12, 13, 13b, 20, 23N}.

A renewed interest in Szekeres spacetimes has appeared, in  relation with  the potential role of density inhomogeneity to explain
the observed cosmic acceleration, without invoking dark energy (see \cite{14, 15, 16, 17, 18, 19, 21, 22} and references therein).

As we shall see below, a heat flux term will appear in the  energy--momentum tensor  of the tilted congruence. As it happens for LTB \cite{38},   such a heat flux vector
would be necessarily associated to a ``truly'' (i.e. entropy producing) dissipative phenomenon, even in the absence of bulk and shear viscosity.

We shall see that as in the nontilted version, the magnetic part of the Weyl tensor vanishes, implying that the super--Poynting vector  asociated to the Bel--Robinson tensor vanishes too, which suggests  the absence of gravitational radiation. Thus confirming the nonradiative character of Szekeres spacetime already stressed in \cite{3}.

Besides the dissipative (irreversible) evolution of the tilted version, and the nongeodesic character of the fluid, another important difference with respect to the nontilted version appears, namely: the vorticity of the fluid lines defined by the tilted four--velocity is now nonvanishing. As it can be expected, this vorticity has important  consequences on the physical interpretation of the model. We shall comment on this issue in the last section.

\section{Szekeres spacetime}
Before considering the tilted version of Szekeres spacetime we find it useful to provide a very brief review of this spacetime in its standard, nontilted, version.

Since the publication of the first two original papers by Szekeres \cite{1, 2} this metric has been presented in many different forms. Here we shall closely follow (with slight changes in notation) the formulation given in \cite{8}.

There are two classes of Szekeres spacetimes, which in Szekeres notation correspond to $\beta^{\prime}\neq 0$ and $\beta^{\prime}= 0$, here we shall deal only with the first class, denoted as LT--type Szekeres metric in \cite{8}.

Thus, the line element is given by:

\begin{equation}
ds^2=-dt^2 + \frac{(R^\prime E-R E^\prime)^2}{E^2 (\epsilon + f)} dr^2 + \frac{R^2 }{E^2}(dp^2+dq^2)
\label{sz}
\end{equation}
where a prime denotes derivative with respect to $r$,  $R=R(t,r)$,  $\epsilon = \pm1,0$ and $f=f(r) > -\epsilon$ is an arbitrary function of $r$. We number the coordinates $x^0=t, x^1=r, x^2= p, x^3=q$.

The function $E$ is given by
\begin{equation}
E(r,p,q)=\frac{S}{2}\left[\left(\frac{p-P}{S}\right)^2 + \left(\frac{q-Q}{S}\right)^2 + \epsilon\right]
\label{E}
\end{equation}
where $S=S(r)$, $P=P(r)$ and $Q=Q(r)$ are arbitrary functions.

From Einstein equations it follows that  $R$ satisfies the equation
\begin{equation}
\dot R^2=\frac{2M}{R}+f,
\label{1}
\end{equation}
where a dot denotes derivative with respect to $t$, and $M=M(r)$ is an arbitrary function. From the above equation it follows that
\begin{equation}
\ddot R=\frac{-M}{R^2},
\label{2}
\end{equation}
from where the meaning of $M$ as an effective gravitational  mass becomes evident.

As in the LTB case there are three subfamilies of solutions depending on the value of $f$, namely; hyperbolic ($f>0$), parabolic ($f=0$) and elliptic ($f<0$).

On the other hand, the parameter $\epsilon$ determines how the  2--surfaces, $r=constant$, foliate the 3D spatial sections of constant $t$. These may be spheres ($\epsilon=+1$), planes ($\epsilon=0$) or quasispheres ($\epsilon=-1$).

For the  ``spherical'' case $\epsilon=+1$, we have the following relationship between coordinates $q, p$ and ``spherical angle'' coordintes ($\theta, \phi$).
\begin{equation}
 \frac{(p-P)}{S}=\cot{\frac{\theta}{2}}\cos{\phi},
 \label{e1a}
 \end{equation}
 \begin{equation}
 \frac{(q-Q)}{S}=\cot{\frac{\theta}{2}}\sin{\phi}.
\label{e1b}
\end{equation}

The LTB spacetime is recovered in the case $E'=0$, $\epsilon=+1$.

For the nontilted (comoving) congruence the source of Szekeres spacetime is pure dust, therefore the energy--momentum tensor is:
\begin{equation}
T_{\mu \nu} = \bar \mu(t,r,p,q) v_\mu v_\nu,
\label{T}
\end{equation}
where vector $v^\mu$ ($v^\mu v_\mu = -1$) is just
\begin{equation}
v^\mu = \left(1, 0, 0, 0\right),
\label{V}
\end{equation}
and $\bar \mu$ denotes the energy momentum density as measured by comoving observers. Then, from the $(t,t)$ component of the Einstein equations we have
\begin{equation}
8\pi \bar \mu=\frac{2(M^{\prime}-3ME^{\prime}/E)}{R^2(R^{\prime}-RE^{\prime}/E)}.
\label{4}
\end{equation}

The two nonvanishing kinematical variables are:

The expansion scalar
\begin{equation}
\bar \Theta=v^\mu_{;\mu}=\frac{\dot R^\prime E - \dot R E^\prime}{R^\prime E-R E^\prime} + 2\frac{\dot R}{R},
\label{theta}
\end{equation}

and the shear  tensor, which can be written as:
\begin{equation}
\bar \sigma_{\mu \nu}=\sigma \left(n_\mu n_\nu - \frac{1}{3}h_{\mu \nu}\right),
\label{shear}
\end{equation}
where
$h_{\mu \nu}= g_{\mu \nu} + v_\mu v_\nu$,  $\bar \sigma^{\mu \nu} \bar \sigma_{\mu \nu}=\frac{2}{3} \sigma^2$  and
\begin{equation}
n^\mu = \left(0, \frac{E \left(\epsilon + f\right)^{1/2}}{R^\prime E-R E^\prime}, 0, 0\right)
\label{S}
\end{equation}
($n^\mu n_\mu = 1$), with
\begin{equation}
\sigma =  \frac{E}{R} \left(\frac{R \dot R^\prime - \dot R R^\prime}{R^\prime E-R E^\prime}\right).
\label{sig}
\end{equation}

Next, the electric part of the Weyl tensor
\begin{equation}
E_{\mu \nu} = C_{\mu \gamma \nu \delta} v^\gamma v^\delta
\label{ewd}
\end{equation}
can be written as
\begin{equation}
E_{\mu \nu} ={ \cal E}\left(n_\mu n_\nu - \frac{1}{3}h_{\mu \nu}\right)
\label{ew}
\end{equation}
where
\begin{equation}
{\cal E}= 3\frac{\ddot R}{R} + 4 \pi \bar \mu.
\label{ecal}
\end{equation}

The magnetic parts of the Weyl and the Riemann tensors vanish identically.

The fact that both the shear tensor and the electric part of the Weyl tensor  have only one nonvanishing independent component, as well as the vanishing of the magnetic part of the Weyl tensor  illustrates how close this spacetime is to the spherical symmetry. This similarity is further reinforced by the fact that the specific forms (\ref{shear}) and (\ref{ew}) are typical of spherical symmetry. Thus the qualification  of ``quasispherical'' assigned by Szekeres himself to his solution \cite{1} looks well justified, in spite of the fact that it has no Killing vectors.

We shall next consider the tilted version of Szekeres spacetime.

\section{Tilted Szekeres spacetime}
In order to obtain a tilted congruence,
let us perform a Lorentz boost from the Locally comoving Minkowskian frame (associated
to $v^\mu$) to  the Locally Minkowskian frame with respect to which a fluid element has ``radial'' (in the $r$ direction)  velocity $\omega$.
For simplicity we shall consider a boost only in the ``radial'' direction and $\omega$ to be a function of $t$ and $r$ alone.

The corresponding tilted congruence is characterized by the four--velocity vector field
\begin{equation}
V^\mu = \left(\frac{1}{(1-\omega^2)^{1/2}}, \frac{E(\epsilon + f)^{1/2}}{R^\prime E - R E^\prime}
\frac{\omega}{(1-\omega^2)^{1/2}}, 0, 0\right).
 \label{Vt}
\end{equation}

For the tilted observer,  the matter distribution described by \eqref{T} will be that given by

\begin{equation}
T_{\mu \nu} = \mu V_\mu V_\nu +  P h_{\mu \nu} + \Pi_{\mu \nu} + q_\mu V_\nu + V_\mu q_\nu
 \label{Tt}
\end{equation}
where $\mu$ is the energy density, $P$ the isotropic pressure, $q^\mu$ the heat flux vector and $\Pi_{\mu \nu}$ the anisotropic pressure tensor, that this observer measures (of course the projector tensor $h_{\mu \nu}$ is now defined in terms of $V^\mu$ instead of $v^\mu$).

The above is the canonical, algebraic decomposition of a second order symmetric tensor with respect to unit timelike vector, which has the above indicated standard physical meaning when $T_{\alpha \beta}$ is the energy-momentum tensor describing some energy distribution, and $V^\mu$ the four-velocity assigned by certain observer.

It is immediate to see that:

\begin{equation} \label{jc10} \mu = T_{\alpha \beta} V^\alpha V^\beta, \; \; q_\alpha = -\mu V_\alpha - T_{\alpha \beta}V^\beta, \end{equation} 
\begin{equation}\label{jc11} P = \frac13 h^{\alpha \beta} T_{\alpha \beta}, \;\; \Pi_{\alpha \beta} = h_\alpha^\mu h_\beta^\nu \left(T_{\mu\nu} - P h_{\mu\nu}\right). \end{equation}

A quick calculation taking into account the above, renders:

\begin{equation}\label{jc2} \mu =\frac{ \bar{\mu}}{1-\omega^2}, \;\; P = \frac{ \bar{\mu}\omega^2}{3(1-\omega^2)},\end{equation}
\begin{equation}\label{jc30} q^\alpha = q N^\alpha, \;\; q = -\frac{\bar{\mu}\omega }{1-\omega^2}, \end{equation}\begin{equation}\label{jc31} N^\alpha  = \left(\frac{\omega}{(1-\omega^2)^{1/2}}, \frac{E(\epsilon + f)^{1/2}}{R^\prime E - R E^\prime}
\frac{1}{(1-\omega^2)^{1/2}}, 0, 0\right),\end{equation}
\begin{equation}\label{jc4} \Pi_{\alpha \beta} = \frac{\bar\mu\omega^2 }{1-\omega^2}(N_\alpha N_\beta-\frac{1}{3}h_{\alpha \beta}), \end{equation}
where use has been made of 
\begin{equation}\label{jc5} v_\alpha = \frac{1 }{(1-\omega^2)^{1/2}} V_\alpha - \frac{\omega}{(1-\omega^2)^{1/2}} N_\alpha.\end{equation}
Alternatively,  it may be useful to define two auxiliary variables ($P_r$ and $P_\perp$) as:  

\begin{equation}
P_r=N^\alpha N^\beta T_{\alpha \beta},\qquad P_\perp=K^\alpha K^\beta T_{\alpha \beta},
\end{equation}
where $K^\alpha$ is a  unit spacelike vector (orthogonal to $V^\alpha$) in the $p$ direction.

Then we may  write

\begin{equation}\label{jc4b} \Pi_{\alpha \beta} = \Pi (N_\alpha N_\beta-\frac{1}{3}h_{\alpha \beta}), \end{equation}
with
\begin{equation}
\Pi=P_r-P_\perp ; \qquad P=\frac{P_r+2 P_\perp}{3},
\label{ns}
\end{equation}
from where the physical meaning of $P_r$ and $P_\perp$ becomes evident.

 Using the above definitions, the  relationships linking tilted and nontilted variables, besides (\ref{jc2}) and (\ref{jc30}), read

\begin{equation}
P_\perp=0; \qquad 3P=\Pi=P_r=\mu \omega^2.
 \label{mu}
\end{equation}

We have next calculated the kinematical variables for the tilted  congruence. These were obtained with MAPLE 13.

The four--acceleration is now nonvanishing, and its components are:

\begin{equation}
a^0=\frac{\omega^2\omega'}{(1-\omega^2)^2}\,\frac{E\sqrt{\epsilon+f}}{R'E-RE'}+\frac{\omega^2}{1-\omega^2}\,\frac{E\dot R'-E'\dot R}{R'E-RE'}+\frac{\omega\dot\omega}{(1-\omega^2)^2},
\end{equation}

\begin{equation}
a^1=\frac{1}{\omega}\,\frac{E\sqrt{\epsilon+f}}{R'E-RE'}\,a^0,
\end{equation}

\begin{equation}
a^2=\frac{\omega^2 \Delta}{1-\omega^2}\,\frac{E}{R}\, \qquad
a^3=\frac{\omega^2 \Gamma}{1-\omega^2}\,\frac{E}{R}\,
\end{equation}
where subscripts $p,q$ denote derivative with respect to these coordinates and 
\begin{equation}
\Delta=\frac{E'_pE-E'E_p}{R'E-RE'}, \qquad\Gamma=\frac{E'_qE-E'E_q}{R'E-RE'}.
\end{equation}

The expansion scalar now becomes
\begin{widetext}
\begin{eqnarray}
\Theta &= &\frac{1}{\sqrt{1-\omega^2}}\,\frac{E R\dot R'+\dot R(2R'E-3RE')}{R(R'E-RE')}+\frac{\omega\dot\omega}{(1-\omega^2)^{3/2}}+\frac{2\omega\sqrt{\epsilon+f}}{\sqrt{1-\omega^2}\,R}+
 \frac{\omega'}{(1-\omega^2)^{3/2}}\,\frac{E\sqrt{\epsilon+f}}{(R'E-RE')}=\nonumber\\
& = & \frac{1}{\sqrt{1-\omega^2}}\,\bar \Theta+\frac{\omega\dot\omega}{(1-\omega^2)^{3/2}}+\frac{2\omega\sqrt{\epsilon+f}}{\sqrt{1-\omega^2}\,R}
+\frac{\omega'}{(1-\omega^2)^{3/2}}\,\frac{E\sqrt{\epsilon+f}}{(R'E-RE')}.
\end{eqnarray}
\end{widetext}

Next, for the vorticity tensor
\begin{equation}
\Omega_{\mu\nu}=\frac{1}{2}\,(V_{\mu;\nu}-V_{\nu;\mu})+\frac{1}{2}\,(a_\mu V_\nu-a_\nu V_\mu)
\label{v1}
\end{equation}
we find the following nonvanishing components
\begin{eqnarray}
\Omega_{02}=\frac{ \omega^2 \Delta}{(1-\omega^2)^{3/2}}\,\frac{R}{E},\qquad \Omega_{03}&=&\frac{\omega^2 \Gamma}{(1-\omega^2)^{3/2}}\,\frac{R}{E},\label{nomega}
\end{eqnarray}
\begin{eqnarray}
\Omega_{12}&=&-\frac{\omega}{(1-\omega^2)^{3/2}}\,\frac{R}{E}\,\sqrt{\epsilon+f}\,(E'_pE-E'E_p),\\
\Omega_{13}&=&-\frac{\omega}{(1-\omega^2)^{3/2}}\,\frac{R}{E}\,\sqrt{\epsilon+f}\,(E'_qE-E'E_q),
\end{eqnarray}

whereas for the vorticity vector:
\begin{equation}
\Omega_\mu=\eta_{\mu\nu\rho\sigma}\,V_{\alpha;\beta}\, V^{\sigma} g^{\nu\alpha} g^{\rho\beta}; \quad \eta_{0123}=-\sqrt{-g},
\label{v2}
\end{equation}
we find:
\begin{equation}
\Omega_0=\Omega_1=0,
\label{V3}
\end{equation}
\begin{eqnarray}
\Omega_2 = \frac{\omega \Gamma}{1-\omega^2}\,\frac{R}{E}\, \qquad
\Omega_3 = \frac{- \omega \Delta}{1-\omega^2}\,\frac{R}{E}. \label{o2}
\end{eqnarray}

Finally, for the shear tensor the following nonvanishing components appear,
\begin{equation}
\sigma_{00}=\frac{2}{3}\,\frac{\omega^2}{1-\omega^2} \sigma_I,
\qquad \sigma_{01}  = -\frac{1}{\omega}\,\sqrt{g_{11}}\,\sigma_{00},
\end{equation}

\begin{equation}
\sigma_{02} = -\frac{1}{2}\,\frac{\omega^2 \Delta}{(1-\omega^2)^{3/2}}\,\frac{R}{E},\qquad
\sigma_{03} =-\frac{1}{2}\,\frac{ \omega^2 \Gamma}{(1-\omega^2)^{3/2}}\,\frac{R}{E},
\end{equation}

\begin{equation}
\sigma_{11} =\frac{1}{\omega^2}\,g_{11}\,\sigma_{00},\qquad
\sigma_{12} = -\frac{1}{\omega}\,\sqrt{g_{11}}\,\sigma_{02},
\end{equation}
\begin{equation}
\sigma_{13} = -\frac{1}{\omega}\,\sqrt{g_{11}}\,\sigma_{03},\qquad
\sigma_{22} = -\frac{1}{2}\,\frac{1-\omega^2}{\omega^2}\,g_{22}  \sigma_{00}, 
\end{equation}
\begin{equation}
\sigma_{33} = \sigma_{22},
\end{equation}
where 

\begin{widetext}
\begin{eqnarray}
\sigma_I= -\frac{\omega}{(1-\omega^2)^{3/2}}\,\frac{\sqrt{\epsilon+f}}{R}+\frac{1}{(1-\omega^2)^{1/2}}\,\frac{E}{R}\,\frac{R\dot R'-R'\dot R}{R'E-RE'}
 +\frac{\omega\dot\omega}{(1-\omega^2)^{3/2}}+\frac{\omega'}{(1-\omega^2)^{3/2}}\,\frac{E\sqrt{\epsilon+f}}{R'E-RE'}.
\end{eqnarray}
\end{widetext}
It is a simple matter to check that the shear tensor may be written as:
\begin{eqnarray}
\sigma_{\alpha \beta}&=&\sigma_I(N_\alpha N_\beta-\frac{1}{3}h_{\alpha \beta})+\sigma_{II}(N_\alpha K_\beta+N_\beta K_\alpha)\nonumber \\
&+&\sigma_{III}(N_\alpha L_\beta+N_\beta  L_\alpha),
\label{sheraa}
\end{eqnarray}
where
\begin{equation}
\sigma_{II}=\frac{\omega \Delta}{2(1-\omega^2)},\qquad
\sigma_{III}=\frac{\omega \Gamma}{2(1-\omega^2)},
\label{osb}
\end{equation}
and $L^\alpha$ is a unit spacelike vectors in the $q$ direction.

Observe that  comparing (\ref{jc4}) with (\ref{sheraa}) it follows that the anisotropy described  by $\Pi_{\alpha \beta}$ cannot be associated exclusively to shear viscosity. 

It is worth stressing that the appearance of vorticity is unavoidable for a generic tilted Szekeres spacetime. Indeed, vanishing of vorticity implies because of (\ref{o2}) 
\begin{equation}
\Delta=\Gamma=0,
\label{nuevar}
\end{equation}
producing
\begin{equation}
\frac{E_p}{E_q}=\psi(p,q),
\label{vor1}
\end{equation}
where $\psi$ is an arbitrary function of its arguments. Then using (\ref{E}) in (\ref{vor1}) we obtain
\begin{equation}
p-P(r)=q\psi(p,q)-Q(r)\psi(p,q)
\label{dm1}
\end{equation}
which implies $P(r)=Q(r)=0$, where the regularity condition $P(0)=Q(0)=0$ has been used. Conditions (\ref{nuevar}) can also be satisfied for some axially symmetric Szkeres models \cite{ln}.

Thus for a generic Szekeres spacetime there is a nonvanishing vorticity according to the tilted observer. Furthermore, as it follows from (\ref{osb}) the shear tensor has the form (\ref{sheraa}), implying three independent components instead of one for the nontilted observer.

It is worth mentioning that the magnetic part of the Weyl tensor vanishes, like for the nontilted case. However the magnetic part of the Riemann tensor 
\begin{equation}
Z_{\mu\nu}=\frac{1}{2}\, \eta_{\mu\rho\alpha\beta}\,R_{\nu\sigma}\, ^{\alpha\beta}\,V^\rho\,V^\sigma
\end{equation}
now has nonvanishing components which are associated to the the dissipative flux $q^\alpha$. They are:
\begin{widetext}
\begin{eqnarray}
Z_{23}=-Z_{32}=-\frac{\omega}{1-\omega^2}\,\frac{R}{E^2}\,\frac{1}{R'E-RE'}\,[E(R\ddot R'+2\ddot R R')+E'(2\dot R^2+R\ddot R)].
\end{eqnarray}
\end{widetext}

Finally, the electric part of the Weyl tensor 
\begin{equation}
E_{\mu\nu}=C_{\mu\alpha\nu\beta}\,V^\alpha\, V^\beta,
\end{equation}
now has the following components:
\begin{eqnarray}
E_{00} = \frac{2}{3}\frac{\omega^2}{1-\omega^2}\,\mathcal{E},\qquad 
E_{01}  =   -\frac{2}{3}\frac{\omega}{1-\omega^2}\,\sqrt{g_{11}}\,\mathcal{E},
\end{eqnarray}
\begin{eqnarray}
E_{22}  =  -\frac{1}{3}\,g_{22}\,\mathcal{E},\qquad
E_{33}  =  E_{22},
\end{eqnarray}
where
\begin{equation}
\mathcal{E}= \,\frac{E}{R}\,\frac{R'\ddot R-R\ddot R'}{R'E-RE'}.
\end{equation}

From the above it follows that we may write the electric part of the Weyl tensor as:
\begin{equation}
E_{\alpha \beta}=\mathcal{E}(N_\alpha N_\beta-\frac{1}{3}h_{\alpha \beta}),
\end{equation}
which is similar to the expression (\ref{ew}) for the nontilted case.

\section{transport equation and entropy production}
In order to evaluate the possibility of entropy production we need a  transport equation for the heat conduction.

Due to the well known pathologies of  Eckart \cite{17T} and Landau \cite{67T} approaches we shall resort to causal dissipative theories. Reasons  for doing that have been extensively discussed in recent years (see \cite{18T, 19T, 20T, 21T, 22T, 23T, 63T, 64T, 65T, 66T, 68T, 69T, 72T, 73T} and references therein).

Thus for a second order phenomenological theory for dissipative fluids we obtain from Gibbs equation and conservation equations (see \cite{64T, 72T} for details):
\begin{widetext}
\begin{eqnarray}
T S^{\alpha}_{;\alpha} =
- q^{\alpha} \left[ h^\mu_{\alpha} (\ln{T })_{,\mu} +
V_{\alpha;\mu} V^\mu
 + \beta_{1} q_{\alpha;\mu} V^\mu+
\frac{T}{2} \left(
\frac{\beta_{1}}{T}V^{\mu}\right)_{;\mu}q_{\alpha}\right],
\label{diventropia}
\end{eqnarray}
\end{widetext}
where $S^\alpha$ is the entropy four--current,  $T$ is temperature and $\beta_1=\frac{\tau}{\kappa T}$, where $\tau$ and  $\kappa$  denote thermal relaxation time and thermal conductivity  coefficient respectively.

Then from the second law of thermodynamics
\begin{equation}
S^\alpha_{;\alpha} \geq 0,
\label{tn1}
\end{equation}
the following transport equation is obtained (see \cite{64T, 72T} for details):
\begin{eqnarray}
\tau h^{\alpha \beta}V^\gamma q_{\beta;\gamma}+q^\alpha&=&-\kappa h^{\alpha \beta}\left(T_{,\beta}+Ta_\beta\right)\nonumber \\&-&\frac{1}{2}\kappa T^2 \left(\frac{\tau V^\beta}{\kappa T^2}\right)_{;\beta} q^\alpha,
\label{tre}
\end{eqnarray}
For simplicity, in the above equations we have not included thermodynamic viscous/heat coupling coefficients, neither have we included couplings of heat flux  to the vorticity. Also, because of the comment below (\ref{osb}) and for simplicity we have  omitted the contributions of shear viscosity ( see \cite{ 64T} for details).

Let us first consider the situation within the context  of the standard (Eckart--Landau) irreversible thermodynamics, in which  case $\tau=0$. Then after simple manipulations  we obtain
\begin{equation}
S^{\alpha}_{;\alpha} = -\frac{1}{2} T^{\alpha \beta}_{dis.}
\mathcal{L}_\chi g_{\alpha \beta},
\label{diventropiaII}
\end{equation}

where $\mathcal{L}_\chi$ denotes the Lie derivative with respect to the vector field $\chi^\alpha=\frac{V^\alpha}{T}$, and  $T^{\alpha \beta}_{dis.}=V^\alpha q^\beta+V^\beta q^\alpha$. From the above it is evident that if  $\chi$ defines  a conformal Killing vector (CKV), i.e.
\begin{equation}
\mathcal{L}_\chi g_{\alpha \beta} =\Phi g_{\alpha \beta},
\label{diventropiaIIIa}
\end{equation}
for an arbitrary function $\Phi$,  then
\begin{equation}
S^{\alpha}_{;\alpha} =0,
\label{diventropiaIII}
\end{equation}
a well known result \cite{35}.

Since  a generic Szekeres spacetime  does not admit CKV \cite{39}, then  the heat flux $q^\alpha$ has to be associated to  an irreversible process.

In the case of the causal thermodynamics the following expression emerges from (\ref{diventropia})

\begin{eqnarray}
S^{\alpha}_{;\alpha} = -\frac{1}{2} T^{\alpha \beta}_{dis.}
\mathcal{L}_\chi g_{\alpha \beta} -\frac{1}{2}\left(\frac{q^2V^\mu \tau}{\kappa T^2}\right)_{;\mu}.
\label{diventropiant}
\end{eqnarray}

Due to the presence of different phenomenological parameters in different terms in (\ref{diventropiant}), it becomes evident that for a generic Szekeres spacetime $S^\alpha_{;\alpha}\neq 0$.

We recall that in the above expression we have neglected terms involving couplings of heat flux  to the vorticity. These couplings give rise to terms of the form $\kappa T \gamma_1 \Omega_{\alpha \beta}q^\beta$,  where $\gamma_1$  is the thermodynamic coupling coefficient, which have to be added to the right hand side of (\ref{tre}). Also, the omitted shear viscosity contribution in (\ref{diventropia}) would give rise to a term of the form $2\eta \gamma_2\pi^\mu_{<\alpha} \omega_{\beta>\mu}$ which has to be added to the transport equation for the shear viscosity, where $\gamma_2$  is a coupling coefficient, $\eta$ denotes the shear viscosity coefficient and $<>$ is the spatial tracefree part of the tensor.

Thus, as in the LTB case \cite{38}, tilted observers in Szekeres spacetime detect a real (entropy producing) dissipative process, while for the nontilted observer the evolution proceeds adiabatically. In this case however it is worth noticing that vorticity contributes  to entropy production  too.

Also, as in the LTB case we might speculate that  the origin of   such important difference in the pictures between both sets of observers may be found in the fact that  while the fluid is geodesic for nontilted observers, it is  not for tilted ones. Then invoking the equivalence  between collisional terms and force terms in the Boltzmann equation established in \cite{74}, it could be possible that the force term  associated  with the nonvanishing four--acceleration plays the role of a collisional term, leading to entropy production.

\section{Conclusions}
We have analyzed the hydrodynamic and thermodynamic properties of the tilted Szekeres spacetime and  compared them with the nontilted  case.

Two main differences emerge from our study: Firstly, the  fluid as seen by the tilted observer evolves irreversibly in contrast with the picture described by the nontilted observers according to which the system evolves adiabatically, furthermore  in the former case the fluid is no longer geodesic. These differences between both set of observers also appear in the FRW and LTB spacetimes.

Secondly and more interesting, the fluid lines as seen by the  tilted observers have nonvanishing vorticity. As mentioned before this is true for any generic Szekeres spacetime.

From a cosmological point of view this might have interesting consecuences. Indeed,  the possibility of rotating universes (besides the well known case of the G\"{o}del  model) and eventual restrictions on the magnitude of the associated vorticity have been studied in the past (see \cite{76, 77, 80, 78} and references therein). Also different mechanisms for vorticity generation in a cosmological  background have been presented (see \cite{79, 82, 81, 83, 75}  and references therein). Here we have seen that the appearance of vorticity may be also an observer related  phenomenon.

We would like to conclude with the following remarks:
\begin{itemize}
\item We have considered the simplest possible case of  tilted congruence, it is obvious that a boost in all possible directions would produce a much   richer  physical picture  from the point of view of tilted observer. In the same line of arguments we have considered the simplest dissipative scenario and have not considered the possible  presence of an electromagnetic field.
The reasons for doing that are on the one hand the cumbersome  resulting expressions and on the other, the fact that even at this level of simplicity  very interesting physical results emerge.

\item  It is worth noticing that, as it follows from (\ref{mu}), if $\bar \mu$ has a good physical behaviour, so do have  the physical tilted variables (at least for small values of $\omega$).

\item In spite of the differences, the tilted version conserves some of the ``structural'' properties of the nontilted one, e.g. the magnetic part of the Weyl tenssor vanishes and the electric part has only one nonvanishing independent component.

\item For a generic Szekeres spacetime with no symmetries and non--barotropic equation of state, it is not obvious that the Gibbbs framework would be applicable (see \cite{nr1}). This point deserves more attention.
\end{itemize}

\begin{acknowledgments}
L.H. thanks  Departamento   de F\'{\i}sica Te\'orica e Historia de la  Ciencia, Universidad del Pa\'{\i}s Vasco and Departament de F\'isica at the  Universitat de les  Illes Balears, for financial support and hospitality. ADP  acknowledges hospitality of the
Departamento   de F\'{\i}sica Te\'orica e Historia de la  Ciencia,
Universidad del Pa\'{\i}s Vasco and Departament de F\'isica at the  Universitat de les  Illes Balears. This work was partially supported by the Spanish Ministry of Science and Innovation (grant FIS2010-15492) and UFI 11/55 program of the Universidad del Pa\'\i s Vasco. JC acknowledges financial support from the Spanish Ministerio de Educaci\'on y Ciencia through grant
no FPA-2007-60220. Partial financial support from the Govern de les Illes Balears is also acknowledged.
\end{acknowledgments}

\end{document}